\documentclass{elsart3p}

 \usepackage{graphicx}

\usepackage{amssymb}

\begin{document}

\begin{frontmatter}



\title{Phase competitions and coexistences in quasi-one-dimensional molecular conductors: 
exact diagonalization study}


\author[label1]{Yuichi Otsuka},
\author[label2]{Hitoshi Seo},
\ead{seo0@spring8.or.jp}
\author[label3]{Yukitoshi Motome},
\author[label4]{Takeo Kato}

\address[label1]{CREST-JST, Department of Material Science, University of Hyogo, Hyogo 678-1297, Japan}
\address[label2]{Synchrotron Radiation Research Center, Japan Atomic Energy Agency,
SPring-8, Hyogo 679-5148, Japan}
\address[label3]{Department of Applied Physics, University of Tokyo, Hongo Bunkyo-ku, Tokyo 113-8656, Japan} 
\address[label4]{Institute for Solid State Physics, University of Tokyo, Kashiwa, Chiba 277-8581, Japan}

\begin{abstract}
We investigate ground state properties of a quasi-one-dimensional electron-lattice coupled model 
for quarter-filled molecular conductors. 
The effective one-dimensional extended Hubbard model 
coupled to adiabatic lattice degree of freedom is derived by the inter-chain mean-field approximation 
and solved  by Lanczos exact diagonalization method. 
We find that the critical behavior among 
lattice tetramerized states with different charge-lattice ordered patterns 
is sensitively affected by the inter-chain Coulomb interaction, 
lattice anharmonicity, and intrinsic dimerization. 
This indicates a subtle balance between these states 
 originating from strong correlation and reduced dimensionality. 
\end{abstract}

\begin{keyword}
Molecular conductors \sep charge order \sep lattice dimerization \sep lattice tetramerization 
\sep spin-Peierls state \sep exact diagonalization 
\PACS 71.10.Fd \sep 71.10.Pm \sep 71.30.+h \sep 71.45.Lr 
\end{keyword}
\end{frontmatter}

\section{Introduction}
\label{intro}
Quasi-one-dimensional (Q1D) molecular conductors exhibit 
a variety of phase transitions, 
where different kinds of symmetry-broken orderings 
in charge, spin, and lattice degrees of freedom
are competing.
Such competitions occur even when their non-interacting band structures are similar, 
suggesting 
a crucial role of interactions, 
i.e., electron-electron as well as electron-lattice interactions. 
Typical examples are the families of (DCNQI)$_{2}X$ and 
(TMTTF)$_{2}X$, which have 
Q1D quarter-filled $\pi$-bands in terms of electrons and holes, respectively~\cite{Review}. 
They show different phase transitions depending on each material 
as well as 
applied pressure,
such as 
charge ordering (CO), 
lattice dimerization, 
which drives the system effectively half-filled 
resulting in a dimer Mott (DM) insulating state, 
and lattice tetramerization 
which is ascribed to a spin-Peierls (SP) transition~\cite{Hiraki-Kanoda-1996,Hiraki-Kanoda-1998,Moret-1988,Chow-2000,Zamborszky,Yu-2004}. 
This diversity originates from an interplay between
the strong Coulomb interaction among electrons 
and the reduced dimensionality giving rise to lattice instabilities. 

Theoretical aspects of such electron-lattice coupled phase transitions 
have been extensively studied 
 based on 1D and Q1D extended Hubbard models
with electron-lattice couplings at quarter-filling~\cite{Seo-Merino-Yoshioka-Ogata-2006}. 
Different states have been reproduced, 
at
the ground state~\cite{Ung-Mazumdar-Toussaint-1994,Riera-Poilblanc-2000,Kuwabara-Seo-Ogata-2003,Clay-2003} 
 as well as 
at
finite temperatures~\cite{Sugiura-2004,Sugiura-2005,QTM,Clay-2007,Otsuka-2008}.
In refs.~\cite{Kuwabara-Seo-Ogata-2003,Clay-2003},
 the ground state phase diagrams 
of 1D models were 
obtained,
 where two different kinds of SP states were found to be competing with each other. 
One is the coexisting state with CO (CO+SP) and 
 the other is that with lattice dimerization, i.e., DM (DM+SP). 
 The phase boundary between these two states is of first order. 
On the other hand, the authors have recently investigated finite-temperature properties of 
 a more general model including 
 the inter-chain Coulomb interaction, a lattice anharmonicity, 
 and the intrinsic dimerization,
and found that they play crucial roles in 
 the critical behavior among different phases, 
 and 
are relevant to 
the experimental systems~\cite{Otsuka-2008}. 
In this paper, motivated by the results, 
 we investigate the effects of such additional contributions 
 on the ground-state phase competitions and coexistences, 
 and how the criticality among different SP states mentioned above are affected. 

\section{Formulation}
\label{form}
We investigate the Q1D quarter-filled model previously studied 
 in refs.~\cite{QTM,Otsuka-2008} at finite temperatures, 
 whose Hamiltonian is given by 
$\mathcal{H} = \sum_{j} \mathcal{H}_{\rm{1D}}^{j} + \mathcal{H}_{\perp}$,
where $\mathcal{H}_{\rm{1D}}^{j}$ and $\mathcal{H}_{\perp}$ are  
the intra-chain part of the $j$-th chain 
and the inter-chain part, respectively.
The former 
is given by 
the extended Peierls-Hubbard Hamiltonian,
\begin{eqnarray}
\hspace{2em} \mathcal{H}_{\rm{1D}}^{j}
 &=& 
- \sum_{i, \sigma} 
\{ t + (-1)^i \delta_{\rm{d}} \}
\left( 1 + u_{i,j} \right) \nonumber\\
&\ & \hspace{5em}  \times \left(
c_{i,j, \sigma}^{\dagger} c_{i+1,j, \sigma} 
+
\rm{h.c.}
\right)
\nonumber \\%
&+&
 \frac{K_{\rm{P}    }}{2} \sum_{i} (u_{i,j})^{2}
  +
 \frac{K_{\rm{P}_{2}}}{2} \sum_{i} (u_{i,j})^{4}
\nonumber \\%
&+&
 U \sum_{i} n_{i,j \uparrow} n_{i,j\downarrow}
 +
 V \sum_{i} n_{i,j} n_{i+1,j},
\end{eqnarray}
where 
the notations are referred to ref.~\cite{Otsuka-2008}.
Here we consider the 
uniaxial lattice distortions $u_{i,j}$ along the 1D chain
and neglect their quantum fluctuations
as in the previous studies~\cite{Kuwabara-Seo-Ogata-2003,Clay-2003}.
The term with $K_{\rm{P}_{2}}$ represents
an anharmonicity in the lattice distortion.
The inter-chain  part is given by
\begin{equation}
 \hspace{3em} \mathcal{H}_{\perp} = V_{\perp} \sum_{\langle j, k \rangle} \sum_{i} n_{i,j} n_{i,k},
\end{equation}
where $V_{\perp}$ denotes the inter-chain Coulomb interaction
and the summation for $\langle j, k \rangle$ runs over nearest-neighbor chains.
In the following calculations, 
we choose the on-site and nearest-neighbor Coulomb interactions
as $U=6$ and $V=2.5$, respectively, 
and the elastic constant as $K_{\rm{P}}=0.75$, 
in the energy unit of $t$~\cite{Otsuka-2008}.

We derive an effective 1D model by the inter-chain mean-field 
 approximation 
 as in ref.~\cite{Otsuka-2008}. 
The mean-field form of $\mathcal{H}_{\perp}$ is taken as%
\begin{eqnarray}
\hspace{2em} \mathcal{H}_{\perp}^{\rm{MF}} = 
\frac{z V_{\perp}}{2} 
\sum_{i} \{
( 
 \langle n_{i-1} \rangle  &+& \langle n_{i+1} \rangle 
) n_{i} \nonumber\\
&-& \langle n_{i-1} \rangle   \langle n_{i+1} \rangle \}, \label{eq:vperp}
\end{eqnarray}%
where $z$ denotes the number of nearest-neighbor chains 
and the chain index $j$ is dropped hereafter.
Eq.~(\ref{eq:vperp}) is derived by assuming 
the relative phase between neighboring chains  
%
as drawn in Fig.~1.
\begin{figure}[t]
\begin{center}
  \includegraphics[width=15em]{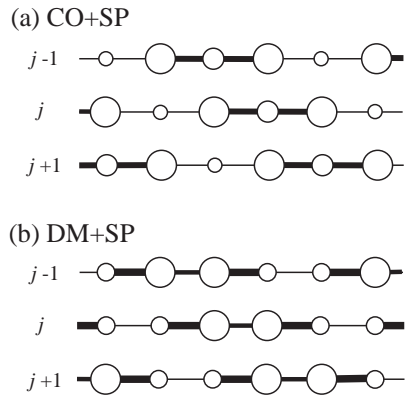}
  \caption{
Schematic view of the interchain configuration between neighboring chains we consider in this work, 
for (a) coexistence of charge order and spin-Peierls (CO+SP) states and 
(b) coexistence of dimerized Mott insulator and spin-Peierls (DM+SP) states. 
Size of the sites and thickness of the bonds represent the charge density and the lattice distortion, 
 respectively. 
The figure is drawn for the two-dimensional case ($z=2$) where 
 chain $j$ is coupled to chains $j-1$ and $j+1$. 
}
 \end{center}
\end{figure}\label{fig1}
We treat $u_i$ adiabatically, 
then static lattice distortions are determined 
so as to minimize the ground-state energy
under the constraint $\sum_{i} u_{i} = 0$.

We obtain 
the ground state of the effective 1D model 
$\mathcal{H}_{\rm{1D}} + \mathcal{H}_{\perp}^{\rm{MF}}$ by exact diagonalization 
using the Lanczos technique, 
and determine $\langle n_i \rangle$ and $u_i$ self-consistently.
We investigate
symmetry breaking with a four-sites period along the chain,
taking into  account all possibility of the CO, DM, and their variants
with further tetramerizations.
The self-consistent equations then involves
the charge densities
$\langle n_{l} \rangle$ and 
the lattice distortions
$u_{l}$ for $l=1,2,3,4$.
In the following, 
we show results for the system size of $L=12$ 
under the anti-periodic boundary condition~\cite{note}.

The obtained self-consistent solutions 
are
parameterized as (see Fig.~1), 
  \begin{eqnarray}
\hspace{2em}   \langle n_{1} \rangle &=& \bar{n}  - n_{\rm CO}  - n_{\rm t (CO+SP)}  - n_{\rm t (DM+SP)}, \nonumber\\
   \langle n_{2} \rangle &=& \bar{n}  + n_{\rm CO}  \quad \quad \quad \quad \quad \ \ 
    + n_{\rm t (DM+SP)}, \nonumber\\
   \langle n_{3} \rangle &=& \bar{n}  - n_{\rm CO}  + n_{\rm t (CO+SP)}  + n_{\rm t (DM+SP)}, \nonumber\\
   \langle n_{4} \rangle &=& \bar{n}  + n_{\rm CO}  \quad \quad \quad \quad \quad \ \ 
  - n_{\rm t (DM+SP)}, \nonumber\\
   u_{1} &=& + u_{\rm d}  - u_{\rm t (CO+SP)},   \nonumber\\
   u_{2} &=& - u_{\rm d}  + u_{\rm t (CO+SP)} + u_{\rm t (DM+SP)},  \nonumber\\
   u_{3} &=& + u_{\rm d}  + u_{\rm t (CO+SP)},   \nonumber\\
   u_{4} &=& - u_{\rm d}  - u_{\rm t (CO+SP)} - u_{\rm t (DM+SP)}, 
 \end{eqnarray}
where $\bar{n}=1/2$ is the average density, 
$n_{\rm CO}$ 
represents the alternating charge disproportionation 
for the CO state, and 
$u_{\rm d}$ the lattice dimerization for the DM state, respectively. 
Lattice tetramerizations described by
$u_{\rm t (CO+SP)}$ and $u_{\rm t (DM+SP)}$
characterize
the CO+SP and the DM+SP states, respectively,
while
four-fold charge disproportionation
concomitantly appears as 
$n_{\rm t (CO+SP)}$ and $n_{\rm t (DM+SP)}$ 
in each of these states.

\section{Results}
\label{result}
First, we show the results
in the absence of intrinsic dimerization, i.e., $\delta_{\rm{d}}=0$. 
Fig.~2 
shows the variation of the order parameters 
 as a function of $zV_\perp$. 
For $zV_\perp < 0.88$ the ground state is 
 the DM+SP state, 
 where \{$u_d$, $u_{\rm t (DM+SP)}$, $n_{\rm t (DM+SP)}$\} are finite 
 while $n_{\rm CO} = n_{\rm t (CO+SP)} = u_{\rm t (CO+SP)} = 0$. 
On the other hand, 
the CO+SP state appears for $zV_\perp > 0.93$  
 where, vise versa, 
 \{$n_{\rm CO}$, $n_{\rm t (CO+SP)}$, $u_{\rm t (CO+SP)}$\} 
 are finite 
 and $u_d = u_{\rm t (DM+SP)} = n_{\rm t (DM+SP)}=0$. 
In the narrow intermediate region, $0.88 < zV_\perp < 0.93$, 
 all of the six parameters 
become
finite. 
This indicates that a further coexistence of 
 the DM+SP and CO+SP states is stabilized, 
 which was not found in previous studies 
 on the ground state properties for similar models~\cite{Kuwabara-Seo-Ogata-2003,Clay-2003}. 
The two phase transition points, 
along 
DM+SP $\leftrightarrow$ the multiple coexistent state $\leftrightarrow$ CO+SP, are of second order. 
\begin{figure}[htbp]
 \begin{center}
  \includegraphics[width=19em]{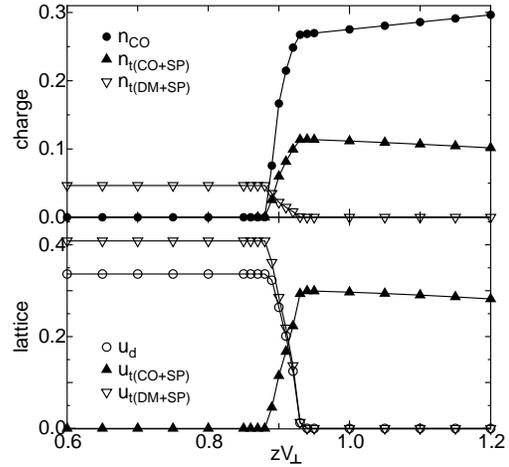}
  \caption{ 
Order parameters as a function of $zV_\perp$ 
for $\delta_{\rm{d}}=0$ for the harmonic case with $K_{\rm P_2}=0$.\label{fig2}}
 \end{center}
\end{figure}

Next, for the case of anharmonic lattice distortions 
with $K_{P_2}=0.75$, 
 the results are shown in Fig.~3. 
Here, we find different critical behavior between 
 the DM+SP and CO+SP states; 
 they are bordered by a direct first order phase transition at $zV_\perp=0.74$. 
This is interesting in a sense that, 
 at finite temperatures above the SP transition temperatures, 
 our recent results show that 
the anharmonicity contributes to rather enlarge
the coexistent region
of the paramagnetic CO and DM states~\cite{Otsuka-2008}. 
\begin{figure}[b]
 \begin{center}
  \includegraphics[width=19em]{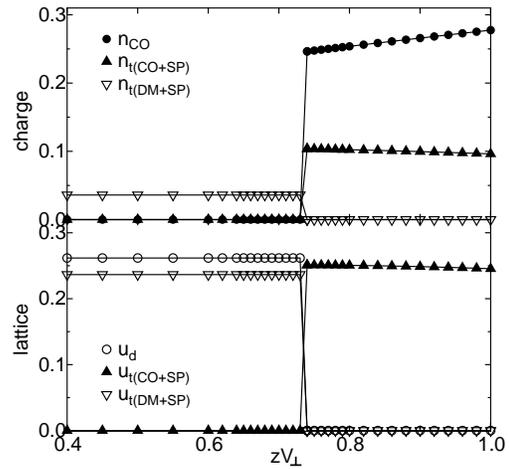}
  \caption{ 
Order parameters as a function of $zV_\perp$ 
for $\delta_{\rm{d}}=0$ for the anharmonic case with $K_{\rm P_2}=0.75$.\label{fig3}}
 \end{center}
\end{figure}

Finally, 
the results in the presence of intrinsic dimerization $\delta_{\rm d}$ 
are shown in Fig.~4; 
here, $u_{\rm d}$ is always finite due to the symmetry breaking from the outset
 in the model. 
For $zV_\perp < 1.0$ the ground state is in the SP(+iDM) phase
 with \{$u_{\rm t (DM+SP)}$, $n_{\rm t (DM+SP)}$\} being finite 
 and $n_{\rm CO} = n_{\rm t (CO+SP)} = u_{\rm t (CO+SP)} = 0$. 
Here iDM refers to the intrinsic dimerization 
 to distinguish with the spontaneous dimerization 
in the model with $\delta_{\rm d}=0$,
 although the actual pattern is the same as in the DM+SP state. 
As for $zV_\perp > 1.0$,
 all the six order parameters become finite; 
 there occurs a second order phase transition into the CO+SP(+iDM) state~\cite{Otsuka-2008}. 
As a result of the intrinsic dimerization leading to $u_{\rm d} \neq 0$, 
 when the CO+SP component arises the order parameters for the SP(+iDM) state 
 always remains, 
 then the charge-lattice pattern is the same as in the multiple coexistence 
 in Fig~\ref{fig2}. 
\begin{figure}[t]
 \begin{center}
  \includegraphics[width=19em]{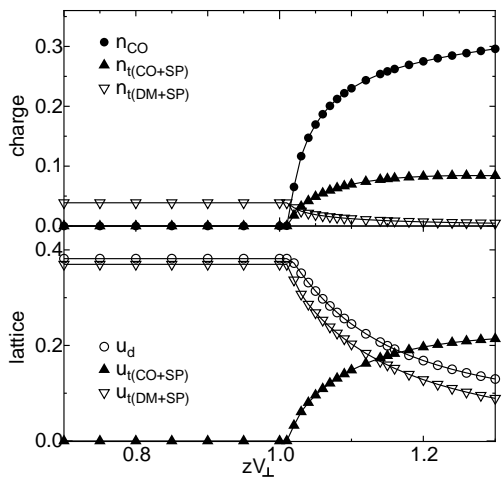}
  \caption{ 
Order parameters as a function of $zV_\perp$ 
 with intrinsic dimerization $\delta_{\rm{d}}=0.02$ 
 for the harmonic case with $K_{\rm P_2}=0$.\label{fig4}}
 \end{center}
\end{figure}

\section{Discussions}
\label{discussion}
Here we compare our results with those in the 
previous studies.
In ref.~\cite{Kuwabara-Seo-Ogata-2003}, 
 the model with $\delta_{\rm d}=0$ and $zV_\perp=0$ was studied. 
There, a first order phase transition between the DM+SP and CO+SP states 
was
found, while the multiple coexistence was not. 
This may be ascribed to the difference 
between
the models, i.e.,
 whether to include $zV_\perp$ or not. 
On the other hand, a model including Holstein-type 
 electron-lattice coupling was considered in ref.~\cite{Clay-2003}, 
 which also indicated the first order phase transition. 
The Holstein coupling under the adiabatic approximation 
leads to a term with a similar but slightly different form 
compared with the inter-chain mean-field contribution
considered in the present work.
This may be the cause of the difference in the results. 
All these comparisons suggest that
the inter-chain Coulomb interaction as well as lattice anharmonicity
affects the critical behavior in the electron-lattice coupled phase transitions.
A comprehensive study including the system size extrapolation is left for future study.

\section{Summary}
\label{summary}
In summary, 
 we have investigated the phase competition and coexistence 
 between different spin-Peierls phases observed in quasi-one-dimensional 
 molecular conductors. 
We found different sequences of phase transitions at the ground state, 
which reflect the slight difference in the model.  
This suggests a subtle balance between 
 charge-lattice coupled symmetry breakings 
 arising from strong correlation and reduced dimensionality. 

\section*{Acknowledgments}
This work is supported by Grants-in-Aid for 
Scientific Research 
(Nos.
18028018, 
18028026, 
19014020)
from the Ministry of Education, Culture, Sports, Science and Technology, and
by Next Generation Integrated Nanoscience Simulation Software.


\begin{thebibliography}{00}
\bibitem{note}
The parameter range we seek for is expected to be where 
the system size dependence is rather small~\cite{Clay-2003,Otsuka-2008}.
\end{thebibliography}
\end{document}